\documentclass[12 pt]{article}\title{Explicit Solution For Klein-Gordon Equation, in Four Dimensions, For any Arbitrary potential. A New Approach }\author{Saeed otarod\thanks{email:
sotarod\@Razi.ac.ir}\\Department of Physics\\ Razi University.
\\Kermanshah, Iran
}\begin{document}\maketitle\baselineskip=2\baselineskip
\begin{abstract}
Klein-Gordon Equation has been solved in  four dimension. The
potential has been chosen to be any arbitrary field Potential.
\end{abstract}
 subject heading:General Physics, Non linear Mathematics, Field Theory.
\cleardoublepage
\section{introduction}
The author in 2002 (1), introduced a new method for solving non
linear Partial differential equations. It seems that this method
is able to solve a vast majority of nonlinear partial Differential
equations. So far, author has been able to solve, Burgurs'
equation (1) and(2), Naviar Stokes equation in one dimension(2)
and Fischer equation. In This artcle we will Solve The well known
Klein - Gordon Equation. As we will see This method makes the
things very simple in many cases.
\section{Solution}
Klein-Gordon Equation in four Dimension Is;
\begin{equation}
\frac{\partial ^{2}}{\partial^{2} t}\phi-\nabla ^{2}\phi+F(\phi)=0
\end{equation}
To solve the problem we take $\phi$ to be a function of another unknown function like $f(x,y,z,t)$
therefore;
\begin{equation}
\phi(x,y,x,t)=\phi(f(x,y,x,t))
\end{equation}
In this way we will have;
\begin{equation}
\frac{\partial\phi}{\partial t}=\frac{d\phi}{df}\frac{\partial f}{\partial t}.
\end{equation}
\begin{equation}
\frac{\partial ^{2}\phi}{\partial^{2} t}=\frac{d^{2}\phi}{df^{2}}(\frac{\partial f}{\partial t})^{2}+\frac{d\phi}{df}\frac{\partial^{2}f}{\partial t^{2}}
\end{equation}
\begin{equation}
\nabla \phi=\frac{d\phi}{df}\nabla f
\end{equation}
\begin{equation}
\nabla ^2\phi=\nabla.\nabla \phi=\frac{d^{2}\phi}{df^{2}}(\nabla {f})^{2}+\frac{d\phi}{df}\nabla ^{2}f
\end{equation}
Substituting equations (4) and (6) in Klein-Gordon equation,,(equation (1)), will lead us to ;
\begin{equation}
\frac{d^{2}\phi}{d f^{2}}[(\frac{\partial f}{\partial t})^{2}-(\nabla f)^2]+
\frac{d\phi}{d f}[\frac{\partial^2 f}{\partial t^{2}}-\nabla ^{2}f]+F(\phi)=0
\end{equation}
Since the function $f(x,y,z,t)$ is quite an arbitrary function we
choose it such that, we can solve the above equation . There may
be several choices, among them we choose $f(x,y,z,t)$ such that;
\begin{equation}
  \nabla f=\overrightarrow{\kappa}=\alpha\hat{i}+\beta\hat{j}+\gamma\hat{k} \hskip 1cm and\hskip 1cm \frac{\partial f}{\partial t}=\lambda
\end{equation}
Here,$ \alpha,\beta,\gamma$ and $\lambda$ are constants. with this
choice, $f$ will come out to be;
\begin{equation}
f(x,y.z.t)=\lambda t+\overrightarrow{\kappa}.\overrightarrow{r}
\end{equation}
here,$\overrightarrow{r}=x\hat{i}+y\hat{j}+z\hat{k}$, and therefore equation(7) will be simplified into;
\begin{equation}
\frac{d^{2}\phi}{df^2}(\lambda^{2}-\overrightarrow{\kappa}^{2})+F(\phi)=0
\end{equation}
To solve this equation we suppose that;
\begin{equation}
\frac{d\phi}{d f}=G(\phi)
\end{equation}
As a result;
\begin{equation}
\frac{d ^{2}\phi}{d f^{2}}=\frac{dG(\phi)}{df}=\frac{dG(\phi)}{d\phi}\frac{d\phi}{d f}=\frac{dG(\phi)}{d\phi}G(\phi)=\frac{1}{2}\frac{d(G(\phi))^{2}}{d\phi}
\end{equation}
Therefore equation(10) will be written as;
\begin{equation}
\frac{1}{2}\frac{d(G(\phi))^{2}}{d\phi}=-F(\phi)
\end{equation}
Separating the variables and integrating both sides of the above equation will give $G$ as a function of $\phi$;
\begin{equation}
G(\phi)=\pm\sqrt{\int-2F(\phi)d\phi}
\end{equation}
Since $G(\phi)$ has been defined to be $G(\phi)=\frac{d \phi}{d f}$, therefore;
\begin{equation}
\frac{d \phi}{d f}=\pm\sqrt{\int-2F(\phi)d\phi}
\end{equation}
\begin{equation}
\int\frac{d \phi}{\sqrt{\int-2F(\phi)d\phi}}=\pm f(x,y,z,t)+\varepsilon
\end{equation}
and using equation(10) we may write the result in a more explicit form ;
\begin{equation}
\int\frac{d \phi}{\sqrt{\int-2F(\phi)d\phi}}=\pm (\lambda t+\overrightarrow{\kappa}.\overline{r})+\varepsilon
\end{equation}
\section{discussion}
Klein-Gordon equation is a well known equation  and has been
investigated by many authors(For example see the references in
reference3). The way we have solved this equation is simple and
straight forward. The result may help us to solve the equation for
any arbitrary potential easily, Of course in cases that the term
in equation(17)is integrable. \\ More over we solved the equation
for a special form of $f(x,y,z,t)$. We have to see what other
choices are possible. If we answer this question, the result will
lead us to other solutions for Klein- Gordon equation that may be,
very important in Physical applications.\\
\section{reference}
1-S.Otarod and J.Ghanbari, Separation of variables for a nonlinear
Differential Equations. Electronic Journal of Differential
Equations, Problem section. 2002-2.\\
2-S.Otarod and J.Ghanbari, Analytical Solution Of Hydrodynamical
Equations. The common project of Razi  and Ferdowsi University.
 Phys.Depts of Razi and Ferdowsi University. Iran. 2003.\\
3-R.Rajaraman., Solitons and Instantons, 1982. Elsevier science
publishing company.\\

\end{document}